# Simulation as Reality? The Effectiveness of LLM-Generated Data in Open-ended Question Assessment

Long (Jim) Zhang[0000-0001-5768-2549] (u3008540@connect.hku.hk)

Faculty of Education, The University of Hong Kong

Meng (Joe) Zhang[0009-0007-6002-9963] (zhangmeng2@kingfa.com)

Kingfa Science and Technology Company, Limited

Wei Lin (William) Wang[0000-0002-1708-1164] (wweilin@connect.hku.hk)

Faculty of Education, The University of Hong Kong

Yu Luo [0009-0001-39652028] (yu.luo3@student.uq.edu.au)

Faculty of Engineering, Architecture and Information Technology, The University of Queensland

**Abstract**

The advancement of Artificial Intelligence (AI) has created opportunities for e-learning, particularly in automated assessment systems that reduce educators' workload and provide timely feedback to students. However, developing effective AI-based assessment tools remains challenging due to the substantial resources required for collecting and annotating real student data. This study investigates the potential and gap of simulative data to address this limitation. Through a two-phase experimental study, we examined the effectiveness and gap of Large Language Model generated synthetic data in training educational assessment systems. Our findings reveal that while simulative data demonstrates promising results in training automated assessment models, outperforming state-of-the-art GPT-4o in most question types, its effectiveness has notable limitations. Specifically, models trained on synthetic data show excellent performance in simulated environment but need progress when applied to real-world scenarios. This performance gap highlights the limitations of only using synthetic data in controlled experimental settings for AI training. The absence of real-world noise and biases, which are also present in over-processed real-world data, contributes to this limitation. We recommend that future development of automated assessment agents and other AI tools should incorporate a mixture of synthetic and real-world data, or introduce more realistic noise and biases patterns, rather than relying solely on synthetic or over-processed data. (207 words)

**1 Main**

The evolution of AI technology has brought generative AI to the forefront, particularly for its capability to produce synthetic simulation data (Cardenuto et al., 2023). Large Language Models (LLMs), notably the Generative Pre-trained Transformers (GPT), have demonstrated remarkable proficiency in generating human-like text (Floridi & Chiriatti, 2020). This advancement has opened new possibilities in synthetic data generation for AI development. A significant breakthrough in this domain comes from Park et al. (2024), who successfully deployed 1,052 generative agents to simulate human responses to social survey questions. Their research yielded compelling results: the generative agents achieved an 85% prediction accuracy, matching the self-consistency of human participants in surveys conducted with a two-week interval. This achievement suggests that synthetic data can effectively substitute for real-world data in training discriminative models, enabling controlled, data-rich experimental environments. These developments hold particular promise for AI tool development, specifically in the context of auto-assessment needs open-ended questions measurement from educational field.

Open-ended questions are particularly significant measurement as such questions can better capture students' understanding and cognitive processes (Birenbaum, 2007). The assessment of open-ended questions involves the evaluation of responses that allow for diverse and subjective answers. For instance, prompts such as "Why is Paris the most memorable place you have ever visited?" require nuanced and subjective interpretation (Çakır & Cengiz, 2016). However, this type of assessment is time-intensive, demands sustained attention, and is prone to scorer subjectivity, posing substantial challenges for automation (del Gobbo et al., 2023). Despite advancements, automating the assessment of open-ended questions remains a significant and complex issue (Urrutia & Araya, 2022). With the surge in online learners, automatic assessment methods are increasingly seen as a means to enhance efficiency in assessment processes while reducing the workload of educators (Viskotová & Hampel, 2022).

The automated assessment of open-ended questions has garnered significant attention in research, with numerous studies proposing diverse methodologies to improve the accuracy and efficiency of such assessments. A critical challenge lies in the inherent subjectivity and complexity of evaluating open-ended responses, which often results in inconsistencies and inaccuracies (Urrutia & Araya, 2022). Traditional automated systems struggle to fully capture the nuances of natural language, necessitating manual intervention to ensure reliability (Nawaz et al., 2022). Commonly employed approaches, such as general evaluation frameworks, rating scales, and rubrics, offer standardized methods for assessment but often fall short in addressing the unique challenges posed by open-ended questions (Koçak, 2020). Some studies have attempted to mitigate these challenges by incorporating crowdsourcing for coding open-ended survey responses, demonstrating an improvement in the quality of evaluations (Jacobson et al., 2017). However, leveraging AI and machine learning has been identified as a more promising direction for overcoming the limitations of traditional approaches in open-ended question assessment. Previous research has established that automatic assessments are a widely recognized and validated approach within e-learning contexts (Amer et al., 2022).

Pre-trained language models, such as DeBERTa (Decoding Enhanced Bidirectional Encoder Representations from Transformers with Disentangled Attention, He et al., 2020), have emerged as powerful tools for tasks requiring textual understanding and classification. DeBERTa, which excels in discriminative tasks like question answering and text classification (Myagmar & Li, 2019), offers stability and precision by relying on pre-existing patterns and knowledge rather than generating content randomly. Fine-tuning is an essential process in adapting pre-trained models like DeBERTa to specific task: assessment of open-ended questions. This process involves further training the model on task-specific labeled data, allowing it to better meet the demands of nuanced evaluations (Swanson, 2024). Specifically, fine-tuning requires a substantial volume of annotated data, where inputs (e.g., questions and responses) are paired with human-evaluated labels. The model iteratively adjusts its parameters based on these labels, improving its ability to predict outcomes accurately. The finetuned DeBERTa have demonstrated promising results in addressing the challenges of open-ended question assessment (Jeong & Kim, 2022; Wang et al., 2022). However, the resource-intensive nature of this process—especially in real-ward, the need for multiple human evaluators to ensure stable and consistent annotations—makes it both costly and time-consuming. As del Gobbo et al. (2023) point out, many studies fail to release their datasets, and the scarcity of available datasets significantly limits the ability to validate and generalize findings. This "dataset crisis" presents a major barrier to the effective use of the DeBERTa model for assessing open-ended questions.

By leveraging Generative LLMs such as GPT to generate large-scale synthetic datasets, DeBERTa can be fine-tuned more efficiently, resulting in a stable and effective automated assessment system for open-ended questions. However, two crucial questions remain unexplored in the current literature: (a) To what extent can LLM-generated synthetic data effectively support open-ended question assessment? (b) What are the practical limitations and boundaries of applying such simulation in real-world educational settings?

Our study employed a rigorous two-phase experimental framework to evaluate the effectiveness of synthetic data as simulation in open-ended question assessment. The framework encompassed comprehensive analysis across four distinct question types, identified through clustering analysis. In Phase I, we developed and evaluated a DeBERTa-based assessment agent trained exclusively on synthetic data. The results demonstrated superior performance of the synthetic data-trained agent in controlled environments. Phase II extended the evaluation to real-world educational settings, where the agent exhibited performance advantages over the state-of-the-art GPT4o model. However, significant performance disparities emerged across different question types and between simulated versus real-world scenarios. This performance gap highlights a fundamental limitation: the controlled experimental environment constructed with synthetic data lacks the inherent noise and biases present in real-world settings. Notably, this limitation from simulation extends beyond synthetic data to heavily processed real-world data, where excessive data processing and feature engineering often eliminate crucial noise patterns through filtering, gap-filling, and outlier removal procedures.

These findings suggest that noise and biases in real-world data may contain essential information for robust AI system performance. We propose that future research should:

1. Preserve and incorporate real-world noise patterns in training data
2. Maintain a balance between data cleanliness and authenticity
3. Consider integrating controlled noise into synthetic data generation

## 2 Experimental Design

This study consists of two phases. In Phase I, our primary objective is to implement an open-ended automated evaluation agent using generated synthetic data. We will discuss this phase in terms of three components: data sources, modeling process, and performance. In Phase II, we validate the performance of this evaluation agent in a real-world educational setting, with GPT-4 as the baseline for comparison. This phase will be examined through two aspects: participants and data collection, as well as data analysis.

### 2.1 Phase I
#### 2.1.1 Data Sources
A public question-answer dataset was used as the test set to evaluate the performance of the evaluation agent. The public dataset is a competition dataset from the Kaggle LLM Science Exam Competition (Will Lifferth, 2023). This dataset contains 200 entries, each in the format of open-ended questions with responses. However, this public dataset was generated by extracting a series of scientific topic text fragments from Wikipedia using GPT-3.5 (for faster running speed, and same with the testing public data in this phase), then asking it to create multiple-choice questions (with known answers), and filtering out simple questions. Below is an example of the format of open-ended questions with responses.

Prompt: What is the butterfly effect?

Responses:

*Simulative student A: The butterfly effect is a physical cause that occurs when a massive sphere is caused to roll down a slope starting from a point of unstable equilibrium, and its velocity is assumed to be caused by the force of gravity accelerating it.*

*Simulative student B: The butterfly effect is a distributed causality that opens up the opportunity to understand the relationship between necessary and sufficient conditions in classical (Newtonian) physics.*

*Simulative student C: The butterfly effect is a proportionality between the cause and the effect of a physical phenomenon in classical (Newtonian) physics.*

*Simulative student D: The butterfly effect is a small push that is needed to set a massive sphere into motion when it is caused to roll down a slope starting from a point of unstable equilibrium.*

*Simulative student E: The butterfly effect is a phenomenon that highlights the difference between the application of the notion of causality in physics and a more general use of causality as represented by Mackie's INUS conditions.*

Best Response by GPT-3.5 generated: E.

Three additional public datasets, along with the generated synthetic data, were used to train the evaluation agent. We utilized the Wikipedia Plaintext dataset (Wikipedia dataset, JJ, 2023) and two STEM-related datasets from Wikipedia (WikiSTEM dataset, Banaei, 2023a; 2023b). The Wikipedia Plaintext dataset (referred to as "Wikipedia" in this study) contains 6,286,775 articles, titles, texts, and categories. The two STEM-related datasets are combined and referred to as "WikiSTEM" in this study, encompassing a wide range of scientific, technical, engineering, and mathematics-related texts. Based on these datasets, we employed the GPT-3.5 model to generate approximately 322,538 entries in the format of open-ended questions with responses as synthetic Data.

**2.1.2 Modeling Process: Fine-tuning DeBERTa Using Synthetic Data**

This model functions as an advanced system for processing and answering open-ended questions by integrating multiple stages. It begins with cleaning and converting data from sources like Wikipedia and WikiSTEM into numerical vectors for efficient similarity searches. These vectors are indexed and stored in a database for quick retrieval of relevant information. The initial query is then enriched with the most relevant retrieved texts, creating a detailed context. The generated simulation 322,538 entries, along with the retrieved context, are used to fine-tune the DeBERTa v3-large model. This fine-tuning process refines the model's accuracy, ensuring the final answers are precise, relevant, and robust.

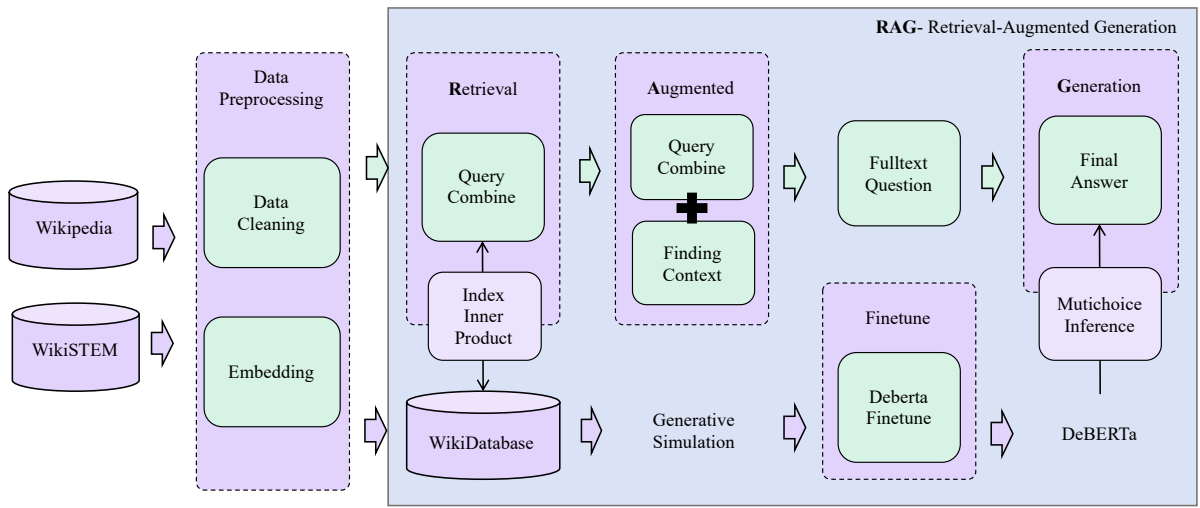

**Figure 1. Utilizing Simulation for Fine-tuning process of DeBERTa**

The model depicted in Figure 1 comprises four main stages: Data Preprocessing, Retrieval, Augmentation, and Generation (RAG, suggested by Siriwardhana et al., 2023), with a crucial Fine-tuning step within the Generation stage to enhance accuracy and performance.

In the Data Preprocessing stage, datasets from open-ended questions, Wikipedia, and WikiSTEM are cleaned and formatted to ensure consistency and quality. After cleaning, the data is embedded into numerical vectors using an embedding model. This embedding process converts textual data into vectors, facilitating efficient similarity searches in the subsequent retrieval stage.

During the Retrieval stage, the embedded vectors are utilized to create indexes for efficient similarity searches. These indexes are stored in the WikiDatabase, which includes data from Wikipedia and WikiSTEM. Tools like FAISS (Facebook AI Similarity Search) are employed to handle these large-scale datasets. The model performs inner product similarity searches through these indexes, retrieving the top 10 most contextually relevant texts. This approach ensures high precision and reduces noise, which is vital for the next stage.

In the Augmentation stage, the initial query is combined with the top retrieved texts to enrich its semantic content. The most relevant contexts from the retrieval stage are selected and integrated with the original query, constructing a comprehensive and contextually accurate augmented query. This augmented query forms a detailed context, which is crucial for generating accurate and relevant answers in the generation stage.

In the Generation stage, the model uses the augmented query and enriched context to generate final answers. The generated simulation 322,538 entries undergo a critical Fine-tuning step, where the DeBERTa model is trained on a dataset of generated questions and answers. This Fine-tuning refines the model's predictive accuracy. Achieved the best performance, we utilized the PyTorch library to implement the model in a Python 3.10 environment on a P100 GPU. Its learning rate was set to 5e-6, a low rate that helps preserve pre-trained features while adapting to new tasks. We used a batch size of 8 for evaluation, and the training spanned 5 epochs. To prevent overfitting, we applied a weight decay of 0.012. Additionally, we utilized a cosine learning rate scheduler with a warm-up ratio and implemented early stopping to enhance training efficiency and maintain model performance. Finally, the fine-tuned DeBERTa model performs multi-choice inference, ensuring that the final answers meet the evaluation criteria.

**2.1.3 Performance Evaluation**

The performance was measured using Mean Average Precision at 3 answers (MAP@3). This metric is widely used in multi-category prediction tasks to assess how well the model can predict the correct answer when multiple possible answers are available. Essentially, MAP@3 evaluates the model's ability to rank the correct answers among the top three predictions. MAP@3 is calculated by averaging the precision scores at each relevant position up to the third rank. The formula is:

$$\text{MAP@3} = \frac{1}{U}\sum_{u=1}^{U}\sum_{k=1}^{\min(n,3)} P(k) \times \text{Rel}(k)$$

In this formula, $U$ represents the total number of questions in the test set; $P(k)$ denotes the precision at the $k$-th prediction; $\text{Rel}(k)$ is an indicator function that equals 1 if the $k$-th prediction is correct (relevant) and 0 otherwise; $n$ is the total number of predicted answers for each question.

We first calculated the performance of the evaluation agent on the test dataset. Furthermore, we categorized this test dataset and computed the performance evaluation for each category to gain deeper insights. For the question classification process, we employed DeBERTa-based text embeddings and K-means clustering algorithms. First, we loaded and preprocessed the 200 scientific questions. To effectively capture the semantic features of each question, we utilized the DeBERTa model to generate embedding vectors. DeBERTa, which stands for Decoding-enhanced BERT with Disentangled Attention, is a state-of-the-art natural language processing model that excels in understanding complex semantic relationships within text (He et al., 2021). By generating high-dimensional vector representations, DeBERTa enables us to capture not only the basic information of the questions but also their underlying semantic similarities.

Following the generation of embedding vectors, we applied the UMAP (Uniform Manifold Approximation and Projection) algorithm in conjunction with K-means clustering. UMAP is a powerful dimensionality reduction technique that preserves both local and global data structures, making it particularly useful for visualizing high-dimensional data (McInnes et al., 2018). This step is crucial as it enables us to visualize the relationships between questions in a more interpretable low-dimensional space. K-means clustering, a widely used algorithm, was employed to partition the questions into distinct clusters based on their semantic similarities. The K-means algorithm is favored for its simplicity and efficiency in handling large datasets, and recent improvements have further enhanced its performance in various applications (Li et al., 2023).

To determine the optimal number of clusters, we implemented the Elbow Method and the Silhouette Score Method. The Elbow Method involves plotting the explained variance against the number of clusters and identifying the point where the rate of variance explained begins to diminish (Thorndike, 1953). The Silhouette Score, on the other hand, measures how similar an object is to its own cluster compared to other clusters, providing a clear indication of the clustering quality (Rousseeuw, 1987). Ultimately, we selected four clusters to ensure a balance between similarity within categories and distinction between categories. This choice facilitates a more nuanced understanding of the semantic relationships among the questions, enabling further analysis and application in various contexts.

Based on the clustering results, we reclassified the questions into four themes: Basic Concepts and Definitions, Advanced Theories and Relationships, Phenomena and Effects, and Practical Applications and Rules. Random sampling of questions from each category was used to verify the reasonableness and accuracy of the classification. Finally, we performed performance evaluations on both the overall dataset and each of the four categories.

### 2.2 Phase II
#### 2.2.1 Participants and Data Collection

We recruited 60 university students from Mainland China for the experiment. They were asked to answer a 10-question open-ended scientific exam. The participants were informed that their participation was completely voluntary, with no rewards or penalties, and no impact on any relevant outcomes, such as course grades. They were required to complete the exam within 60 minutes. This ensured that the participants could respond with confidence, thus guaranteeing the validity of the data.

The 60 participants were randomly divided into 12 groups, with 5 participants in each group. The answers from each group were randomly drawn from one of the four sets of question-answer papers we provided: Basic Concepts and Definitions, Advanced Theories and Relationships, Phenomena and Effects, and Practical Applications and Rules. These papers covered the same range of knowledge but used questions defined from different perspectives. By using this diverse questioning approach, we ensured the representativeness of the data.

Additionally, we recruited three independent science teachers with teaching experience to evaluate the students' responses. Each teacher was tasked with identifying the best response (Top 1) and the relatively higher-level responses (Top 3) for each of the five answers within each question. The evaluation results from the three teachers were combined using the following formula to represent the overall ranking of the responses. The complete representation using $A\_i$ where i = 1, 2, 3, 4, 5 is:

$$\text{Score}(A\_i) = (7 \times H\_i + 3 \times M\_i + 1 \times L\_i) / 3$$
$$\text{Order} = \text{sort}(\text{Score}(A\_i), \text{descending})$$

$H\_i$ represents responses that some teachers consider to be Top 1; $M\_i$ represents responses that some teachers consider to be Top 3 but not Top 1; $L\_i$ represents responses that are not in the Top 3. Ultimately, each question-and-answer pair and the teacher's annotations are organized into the following example format.

Prompt: What are redox reactions?

Responses:

*Student A: A redox reaction is any reaction in which there is a rise or fall in the valence of an element. The essence of a redox reaction is the gain or loss of electrons or the shift of shared electron pairs. Redox reactions are one of the three basic reactions in chemistry*

*Student B: Redox reactions are a class of reactions in which there is a change in the oxidation number of an element before and after a chemical reaction. The essence of a redox reaction is the gain or loss of electrons or the shifting of shared electron pairs. Redox reactions are one of the three basic reactions in chemistry (the other two are (Lewis) acid-base reactions and free radical reactions). Combustion, respiration, photosynthesis in nature, chemical batteries, metal smelting, rocket launching*

*and so on are all closely related to redox reactions. The study of redox reactions is of great importance to the progress of mankind.*

*Student C: Redox reactions are chemical reactions in which elements or groups in a substance undergo electron transfer before and after the reaction.*

*Student D: Redox reactions are a class of reactions in which there is a change in the oxidation number of an element before and after a chemical reaction. The essence of a redox reaction is the gain or loss of electrons or the shifting of shared electron pairs.*

*Student E: A chemical reaction in which the valence of an element rises or falls during the reaction is a redox reaction.*

Response Order by three teachers: D, B, A, C, E.

**2.2.2 Data Analysis**

To comprehensively evaluate the performance of this agent in real educational settings, we defined three sub-goals: identifying the best response (Top1), determining whether the response is among the top three (Top3), and evaluating overall ranking. For each of these sub-goals, we conducted both holistic assessments and evaluations on four distinct types of problem sets.

The Top1 task essentially aims to assess the accuracy of the model's prediction of the highest-ranked option. Specifically, it checks whether the model's top-ranked choice matches the human-assigned first place. This is a typical classification task where we focus solely on whether the model correctly predicts the most important option, disregarding the order of the remaining rankings. To evaluate performance on this "first place prediction" task, we used accuracy, precision, recall, and F1 score.

The Top3 task, in essence, is a binary classification problem. Each option is categorized as either a "good answer" (within the top three) or "other answers" (outside the top three). The model's task is to determine whether each option should be classified as a "good answer." This evaluation does not concern itself with the specific order of the top three responses, but instead assesses the model's ability to identify relatively high-quality options.

For both the Top1 and Top3 tasks, performance was measured using accuracy (Acc), precision, recall, and F1 score. Accuracy reflects the proportion of true results among the total cases examined, providing an overall measure of how frequently the model correctly classifies responses. Precision (or positive predictive value) is the ratio of correctly predicted positive outcomes to the total predicted positives, indicating the reliability of the model in identifying high- or medium-quality responses. Recall (or sensitivity) is the ratio of correctly predicted positive outcomes to all actual positive cases, measuring the model's ability to capture all actual high/medium-quality responses. F1 score, the harmonic mean of precision and recall, balances the trade-off between these two metrics, offering a single measure that accounts for both false positives and false negatives. This score is particularly useful in scenarios with imbalanced class distributions.

$$\text{Acc} = \frac{TP + TN}{TP + YN + FP + FN}$$

$$\text{Precision} = \frac{TP}{TP + FP}$$

$$\text{Recall} = \frac{TP}{TP + FN}$$

$$F1 = 2 \times \frac{\text{Precision} \times \text{Recall}}{\text{Precision} + \text{Recall}}$$

Where TP is True Positives. TN is True Negatives. FP is False Positives. FN is False Negatives. These metrics help in comprehensively evaluating the model's performance in classifying the responses into higher- levels, providing insights into both the correctness and completeness of the model's predictions.

For the overall ranking task, we evaluated the consistency between the large language model's performance and human expert judgments in a multi-option ranking task. Specifically, the model was required to rank the given options based on their importance or relevance, and these rankings were compared with those provided by human experts. We employed three complementary evaluation metrics: NDCG, which assesses the accuracy of the overall ranking by considering the positions of all relevant items, thereby providing an overall quality measure of the ranking; and MAP, which also evaluates the ranking accuracy by taking into account the positions of all correct items, offering a comprehensive evaluation of the ranking's quality. This multidimensional evaluation approach allows for a thorough understanding of the model's performance in ranking tasks, providing reliable metrics to guide both the comprehension and enhancement of the model's ranking capabilities.

$$\text{NDCG} = \frac{\text{DCG}}{\text{IDCG}} = \frac{\sum_{i=1}^{K} \frac{2^{\text{rel}_i} - 1}{\log_2 (i + 1)}}{\sum_{i=1}^{K} \frac{2^{\text{rel}_i^*} - 1}{\log_2 (i + 1)}}$$

In this context, $\text{rel}_i$ represents the relevance score of the item at the i-th position, while K is the length of the ranking list, or the specified evaluation cutoff position. IDCG stands for the ideal DCG, which is the DCG when the relevance scores are sorted in descending order, meaning $\text{rel}_i^*$ is the relevance score of the i-th position after sorting the items by their true relevance scores from high to low. The range of NDCG is between 0 and 1, where a value of 1 indicates a perfectly predicted ranking.

$$\text{MAP} = \frac{1}{Q} \sum_{q=1}^{Q} \text{AP}_q = \frac{1}{Q} \sum_{q=1}^{Q} \frac{\sum_{k=1}^{n} P(k) \cdot \text{rel}_i}{n}$$

In this context, Q refers to the query set, AP represents the Average Precision, P(k) is the precision at position k, and rel(k) indicates whether the item at position k is relevant (0 for irrelevant, 1 for relevant). m is the total number of relevant items, and **n** is the length of the ranked list. The value of MAP (Mean Average Precision) ranges from 0 to 1, where a value of 0 indicates that the model has not returned any relevant documents correctly for any of the queries, while a value of 1 means that the model has correctly returned all relevant documents and the ranking is perfectly accurate across all queries.

## 3 Result
### 3.1 Phase I

When clustering the questions from 200 publicly available datasets, both the elbow method and silhouette coefficient method suggest that four clusters should be chosen. In the elbow method plot, the SSE (Sum of Squared Errors) begins to level off noticeably at four clusters, forming an "elbow," which indicates that four clusters is a reasonable choice. In the silhouette coefficient plot, the silhouette score reaches a relatively high value (greater than 0.5) at four clusters, indicating that the clustering is effective. Thus, selecting four clusters is appropriate.

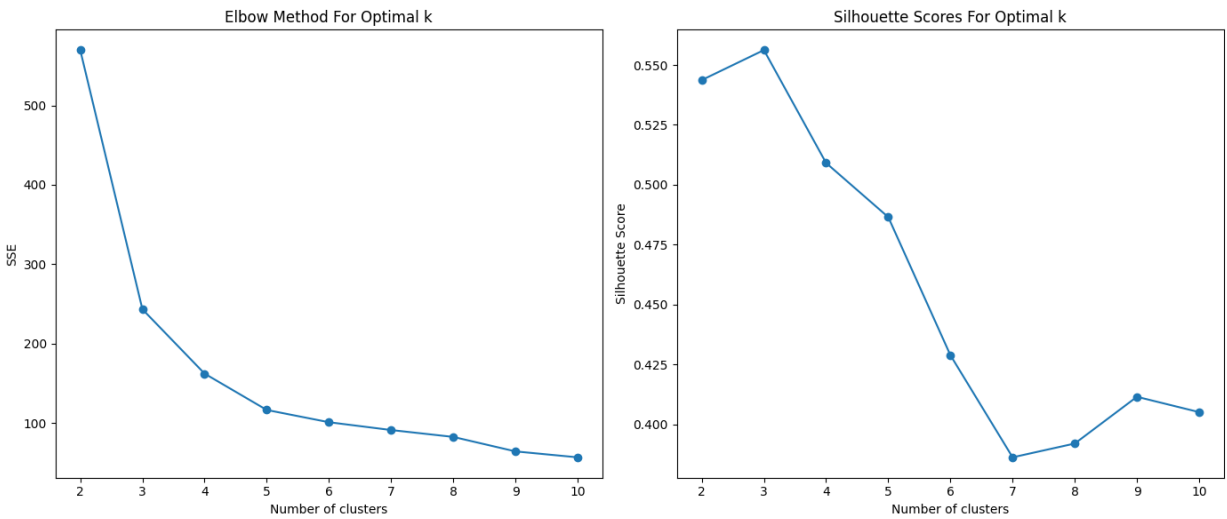

**Figure 1 The Elbow Map & Silhouette Score Map with UMAP + K-Means**

Ultimately, we successfully divided the data into four clusters, with the largest cluster containing 68 samples and the smallest containing 37, showing a fairly balanced sample distribution. The clustering performance was evaluated using three key metrics: the silhouette coefficient, which reached 0.509, indicating good cohesion within each cluster and appropriate separation between clusters; the Calinski-Harabasz index, which was 659.867, reflecting a high level of tightness within clusters and differentiation between clusters; and the Davies-Bouldin index, which was 0.660, further confirming good separation between clusters. Therefore, the clustering was successful, offering strong interpretability and practical value.

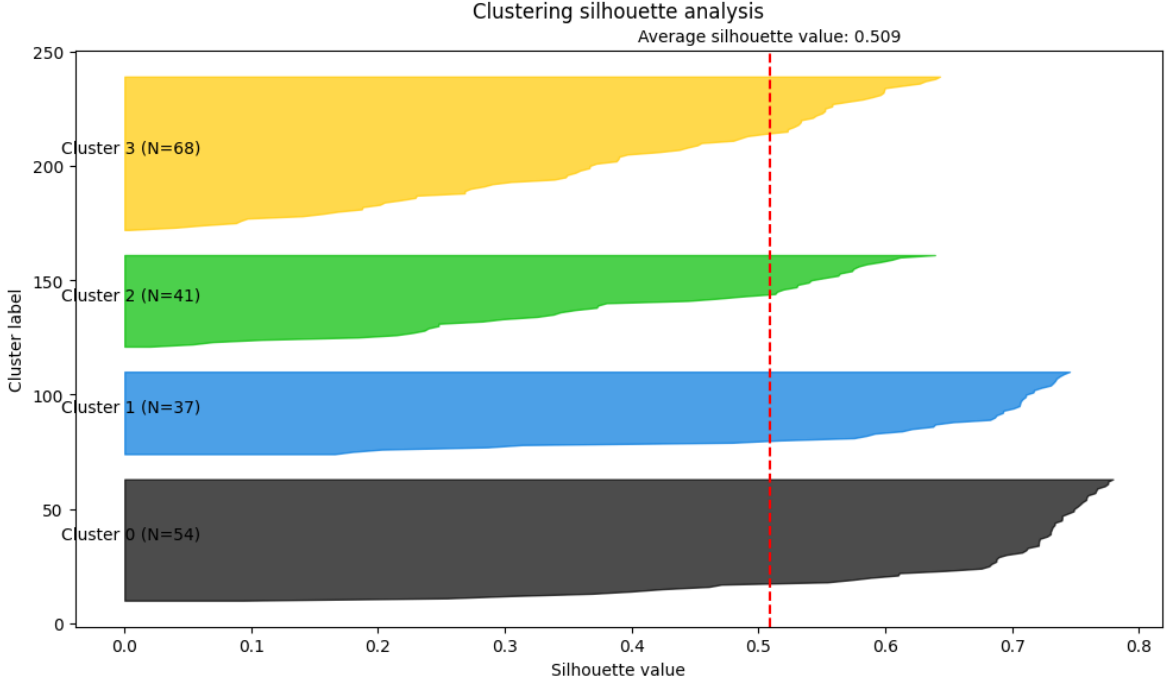

**Figure 3**

We categorized these questions into four classes based on key feature words. The specific key features for each category are shown in Table 2. Concepts and Definitions: Help students grasp fundamental definitions and concepts. Theories and Relationships: Require students to understand and explain more complex theories and relationships. Phenomena and Effects: Focus on specific phenomena and effects and their scientific explanations. Applications and Rules: Emphasize practical applications and operational rules, helping students understand how to apply theoretical knowledge in practice.

**Table 1 Questions Clusters**

| Cluster | Top-keywords | Explanation | Representation |
| --- | --- | --- | --- |
| Cluster 0: Concepts and Definitions | Definitions, Effect | This theme includes questions that deal with basic definitions and fundamental concepts that form the foundation of a subject. It often involves introductory terms and descriptions of key principles. | What is the Roche limit? |
| Cluster 1: Theories and Relationships | Used, Relationship, Earth | This theme encompasses questions that require an understanding of advanced theories and complex relationships within the subject matter. It includes detailed explanations of intricate principles that go beyond basic understanding. | What is the revised view of the atmosphere's nature based on the time-varying multistability that is associated with the modulation of large-scale processes and aggregated feedback of small-scale processes? |
| Cluster 2: Phenomena and Effects | Relationship, Role, Following | Questions under this theme involve specific phenomena and effects that are observable or measurable. This category includes questions that explain how certain phenomena occur and the effects they produce. | What is the 'reactive Leidenfrost effect' observed in non-volatile materials? |
| Cluster 3: Applications and Rules | Physics, Main, Reason | This theme is focused on the practical application of theories and concepts. It includes questions about how theoretical knowledge is applied in real-world scenarios, including rules and procedures that govern practical applications. | What is the role of axioms in a formal theory? |

After determining the clusters, we evaluated the model's overall performance as well as its performance on each category. Ultimately, we analyzed the incorrect answer options generated by the model, which achieved a MAP@3 local test score of 0.936. Notably, the MAP@3 score of 0.532 was obtained when using DeBERTa alone without the enhancement from our method. This highlights the success of our proposed method in improving DeBERTa's performance on this open-ended question evaluation task.

Given the high MAP@3 score overall, we focused on the 22 question-answer entries where the best answer was not successfully predicted. Among these 22 entries, 17 had their correct answers predicted in the second position. The other 5 were predicted in third or lower positions, which we focused on. These mispredictions were mainly concentrated in the "Applications and Rules" category (3/5), while "Theories and Relationships" and "Phenomena and Effects" each had one misprediction (2/5). This suggests that the model may perform less well in predicting the best answer for question entries in the "Applications and Rules" category compared to the other categories. This evaluation was conducted on a virtual dataset, and the following study will report on the model's performance in a real educational environment.

## 3.2 Phase II

**Table 2**

| Question Type | Task | Model | Accuracy | Precision | Recall | F1 |
|---|---|---|---|---|---|---|
| Overall | Top1 | Ours | **0.442** | 0.412 | **0.416** | **0.412** |
| | | GPT4o | 0.408 | **0.429** | 0.407 | 0.395 |
| | Top3 | Ours | **0.575** | **0.731** | **0.731** | **0.731** |
| | | GPT4o | 0.545 | 0.706 | 0.706 | 0.706 |
| Applications and Rules | Top1 | Ours | **0.533** | **0.566** | **0.573** | **0.53** |
| | | GPT4o | 0.433 | 0.372 | 0.360 | 0.357 |
| | Top3 | Ours | **0.636** | **0.778** | **0.778** | **0.778** |
| | | GPT4o | 0.579 | 0.733 | 0.733 | 0.733 |
| Concepts and Definitions | Top1 | Ours | **0.400** | 0.311 | **0.310** | **0.295** |
| | | GPT4o | 0.375 | **0.319** | 0.295 | 0.276 |
| | Top3 | Ours | 0.519 | 0.683 | 0.683 | 0.683 |
| | | GPT4o | **0.529** | **0.692** | **0.692** | **0.692** |
| Phenomena and Effects | Top1 | Ours | **0.450** | **0.270** | **0.338** | **0.295** |
| | | GPT4o | 0.250 | 0.165 | 0.252 | 0.186 |
| | Top3 | Ours | **0.500** | **0.667** | **0.667** | **0.667** |
| | | GPT4o | 0.463 | 0.633 | 0.633 | 0.633 |
| Theories and Relationships | Top1 | Ours | 0.400 | 0.288 | 0.421 | 0.302 |
| | | GPT4o | **0.533** | **0.412** | **0.488** | **0.427** |
| | Top3 | Ours | **0.651** | **0.789** | **0.789** | **0.789** |
| | | GPT4o | 0.593 | 0.744 | 0.744 | 0.744 |

Table 2 reports the evaluation performance comparison between our agent model and GPT4o across overall and four types of questions. Our model demonstrates superior performance on both the Top1 (identifying the best option) and Top3 (ranking options in terms of high and low levels) core tasks. In the Top1 task, which assesses the accuracy of predicting the best option, our model achieves an overall accuracy of 44.2%, significantly outperforming GPT4o's 40.8%. This indicates that our model has stronger judgment ability in identifying the most important option. Notably, in the "Applications and Rules" category, our model performs particularly well in the Top1 task, with an accuracy of 53.3%, far exceeding GPT4o's 43.3%, which demonstrates the model's advantage in practical application scenarios.

In the Top3 task, which is a binary classification problem of categorizing options into "Top Three" and "Others," our model also shows excellent performance. The overall F1 score reaches 0.731, surpassing GPT4o's 0.706, indicating that our model is more accurate in identifying high-quality options. It is worth noting that in this task, our model demonstrates more balanced performance, with both precision and recall reaching 0.731, indicating that the model maintains high accuracy while not missing important options. Particularly in the "Applications and Rules" and "Phenomena and Effects" categories, our model's F1 scores in the Top3 task (0.778 and 0.667, respectively) are significantly higher than GPT4o's, proving the model's stronger judgment ability in recognizing multiple reasonable options.

**Table 3**

| Question Type | Model | NDCG | MAP |
|---|---|---|---|
| Overall | Ours | **0.920** | **0.660** |
| | GPT4o | 0.914 | 0.627 |
| Applications and Rules Quiz | Ours | **0.932** | **0.702** |
| | GPT4o | 0.926 | 0.659 |
| Concepts and Definitions Quiz | Ours | **0.917** | **0.640** |
| | GPT4o | 0.903 | 0.605 |
| Phenomena and Effects Quiz | Ours | **0.893** | **0.631** |
| | GPT4o | 0.873 | 0.514 |
| Theories and Relationships | Ours | 0.932 | 0.662 |
| | GPT4o | **0.946** | **0.698** |

In the overall ranking task, our evaluation agent also holds a clear advantage. From a general performance perspective, our model achieves better results on two key metrics, NDCG and MAP (NDCG: 0.920 vs 0.914, MAP: 0.660 vs 0.627), indicating that our model excels in both ranking quality and retrieval accuracy. Notably, in the "Applications and Rules Test" and "Concepts and Definitions Test"—two scenarios that require deep understanding—our model performs exceptionally well, with NDCG values of 0.932 and 0.917, respectively, showing significant improvements over GPT4o.

When handling "Phenomena and Effects Test" type questions, our model demonstrates greater robustness. The notable improvement in the MAP metric (0.631 vs 0.514) suggests that the model exhibits better judgment ability when dealing with complex phenomenon explanations. Although GPT4o slightly outperforms our model in the NDCG metric for "Theories and Relationships" type questions (0.946 vs 0.932), our model still maintains a high level of performance, indicating that it remains competitive when addressing abstract theoretical relationships.

## 4 Discussion

This study demonstrates the potential of enhancing open-ended question assessment through a DeBERTa-based agent trained on 322,538 synthetic data points. Our findings reveal two significant outcomes. First, in Phase I, the DeBERTa model successfully achieved fine-tuning optimization using synthetic data, showing promising performance in controlled environments. Second, Phase II demonstrated that the assessment agent can maintain effectiveness in real-world applications, generally outperforming the GPT4o benchmark model. However, a notable performance disparity emerged between synthetic and real-world scenarios. While the agent exhibited superior performance in synthetic data testing (Phase I), its real-world application (Phase II) revealed room for improvement. Based on these findings, we propose a guiding framework that advocates for the integrated use of synthetic and real-world data in AI model development. This hybrid approach could potentially maximize the utility of simulation in training and fine-tuning AI tools while maintaining real-world effectiveness.

### 4.1 Success of fine-tune assessment agent by simulation

In Phase I, our synthetic data-based fine-tuning of the DeBERTa assessment agent yielded remarkable results. When tested against synthetic datasets, the agent demonstrated exceptional performance, achieving a MAP@3 score of 0.935 - a dramatic improvement from the baseline of 0.532. While these results were promising, a crucial question remained: Would this success translate to real-world environments? The answer was affirmative as hypothesized.

In Phase II, we deployed the fine-tuned assessment agent in real-world educational settings, testing it against real student responses and teacher assessments. To our delight, the agent maintained its effectiveness, and even generally outperformed GPT4o across performance. This success becomes particularly significant considering our agent's practical advantages - it avoids generated LLM's hallucination issues and resource-intensive nature (Cf. Hadi et al., 2024), making it more suitable for open-ended auto-assessment tasks that demand both reliability and speed.

From above, this study also supports the findings of that the synthetic data benefit to fine-tune and train AI tools' development (e,g, Kortylewski et al., 2019; Kouam et ., 2024). In our research, the synthetic data we used guided the training of the autonomous evaluation agent and was successfully implemented in real-world teaching scenarios. This indicates that while there may be a gap between simulated and real data, both can be used to guide the fine-tuning of LLMs. Simulation such as synthetic data is typically generated through specific algorithms or models designed to replicate certain features or behaviors of the real world (Cf. Nassif et al., 2024). The generation process allows for controlled variables, ensuring data diversity and coverage, thus providing rich samples for model training (Cf. Paproki et al., 2024). For example, in open-ended question evaluations, synthetic data can be designed to include various types of questions and difficulty levels, enabling the model to perform across different scenarios.

**4.2 Gap and implication from simulation data using**

Despite our assessment agent's success in outperforming GPT4o, our analysis reveals a noteworthy performance disparity between simulative and real-world testing. The simulation-trained assessment agent consistently achieved superior results in simulative environments by synthetic, yet showed measurable gaps when compared to human assessments in real-world applications. This performance differential likely stems from our training approach - the agent, trained exclusively on synthetic data, naturally excels in similar synthetic contexts. This observation highlights a critical caveat in AI tools development: while simulation, such as synthetic data, can effectively bootstrap AI tools in controlled environments, it may introduce inherent limitations when deployed in real-world scenarios.

This limitation is that it may not fully capture the complexity and randomness of the real world. Real-world data often contains more noise, outliers, and uncertainties (Hariri et al., 2019), which might be overlooked or simplified in synthetic data. Synthetic data helps the model quickly learn specific patterns and rules, but due to the limitations in its generation process, the model might show bias when faced with real-world data. This bias stems from differences between the sample characteristics encountered during simulation training and the actual application in real-world scenarios. For instance, while models may effectively learn response patterns from synthetic data, they often struggle when confronted with the inherent diversity and complexity of real-world questions, resulting in compromised assessment accuracy. This limitation extends beyond synthetic data to encompass processed real-world data as well. Notably, conventional data preprocessing and feature engineering practices - including missing value imputation, noise filtering, and outlier removal (e.g., Calabrese, 2018) - may inadvertently eliminate crucial contextual information. Just like "over-sanitization" of training data, whether synthetic or real, presents a paradox in AI development: the very processes intended to enhance model performance may actually limit its ability to handle real-world variability. The standardized patterns learned from cleaned and synthetic data might fail to capture the nuanced characteristics that define authentic educational assessment scenarios.

Based on our findings, we caution against exclusive reliance on over-idealized data and training environment - whether synthetic or heavily processed - for AI tool development and evaluation. While such sanitized data provides a controlled training environment and may reduce task-specific errors, it risks creating a disconnect between AI performance and real-world applications. Although controlled environments demonstrate lower error rates, this apparent success often fails to translate to real-world scenarios, where complexity and unpredictability introduce higher error probabilities.

The uncertainty and variability inherent in real-world data represent not merely noise to be eliminated, but essential characteristics that AI systems should learn to navigate (Cf., Crusius et al., 2025). The current prevalent practice in AI research - utilizing synthetic or highly processed "clean" data for training and testing - may create an illusion of excellence that poorly predicts actual performance. The disconnects between simulation and reality potentially stems from the exclusive use of idealized training data.

As this study highlights the critical importance of preserving real-world data characteristics during model training to ensure reliable performance in actual applications. We therefore propose a more nuanced approach to data quality in AI development. Firstly, while clean, well-annotated synthetic and processed data remain important, training data should intentionally incorporate real-world noise and biases. Secondly, rather than pursuing maximum data cleanliness, researchers should consider a spectrum of data quality that reflects real-world complexity. Thirdly, future research should explore the potential information embedded within noise and biases, as these patterns may enhance data quality and model robustness. Moreover, a promising direction for future research lies in investigating the deeper patterns within apparent noise and biases these insights into training data development.

Looking forward, we advocate for a synergistic approach that integrates data from simulated controlled (including synthetic or processed data) and real-world environment in AI development. Rather than viewing these data types as opposing forces, we recognize their complementary nature in enhancing AI capabilities. Synthetic data provides an invaluable foundation for initial model training and task-specific fine-tuning, while real-world data offers essential contextual validation and environmental adaptation signals (e.g. Kortylewski, 2019). This complementary relationship suggests an iterative development pathway where models can continuously evolve and optimize their performance.

We suggested a progressive training paradigm where initial synthetic data bootstraps the model's foundational capabilities, followed by gradual exposure to real-world interactions. As the AI system scales in deployment, it can accumulate authentic interaction data, enabling continuous refinement through real-world feedback loops. This evolutionary approach allows the model to maintain the advantages of controlled synthetic training while progressively adapting to the nuances and complexities of real-world applications. Such a dynamic training strategy could potentially bridge the synthetic-real performance gap while leveraging the strengths of both data types.

## 5 Conclusions

In this study, we provide controlled simulation experimental environment LLM-generated synthetic data by to Implement an open problem assessment AI tool agent. It achieved a successful fine-tuning of DeBERTa agent with MAP@3 score of 0.936 on a public testing dataset, which is simulation dataset. Then, to further demonstrate the validity of this assessment model by simulation, we performed the experiment in a real teaching scenario. Our empirical findings demonstrate that our model achieves superior evaluation performance compared to GPT4o (the state-of-the-art model at the time of writing), despite operating at a significantly smaller scale. However, the persistent gap between our model's assessments and human evaluations underscores a fundamental challenge: AI tools trained and tested primarily in simulated environments with synthetic or heavily sanitized data may face inherent limitations in real-world applications. This observation reinforces our recommendation for an integrated approach that values both synthetic and real-world data, including the inherent noise and variations in authentic data.

While our study offers meaningful insights, several limitations warrant acknowledgment. First, our investigation focused solely on open-ended question assessment and benchmarked against a single Generative LLM (GPT4o), suggesting opportunities for broader comparative studies. Second, space constraints prevented a thorough examination of how different question types might influence AI tool capabilities. Third, our focus on science-domain questions leaves room for expansion into other academic disciplines. Finally, the scope of our synthetic data and participant pool could be expanded to enhance the robustness of our findings.

(6608 words)